\documentclass[doublecol]{epl2} 
\usepackage{graphicx}
\usepackage{epsfig}
\usepackage{epstopdf}
\title{The melting of the classical two dimensional Wigner crystal}
\shorttitle{Wigner crystal} 
\author{M. Mazars\inst{1}}
\shortauthor{M. Mazars}
\institute{\inst{1} Laboratoire de Physique Th\'eorique (UMR 8627), Universit\'e de Paris-Sud and CNRS, B\^atiment 210, 91405 Orsay Cedex, FRANCE }

\abstract{We report an extensive Monte-Carlo study of the melting of the classical two dimensional Wigner crystal for a system of point particles interacting via the $1/r$-Coulomb potential. A hexatic phase is found in systems large enough. With the multiple histograms method and the finite size scaling theory, we show that the fluid/hexatic phase transition is weakly first order. No set of critical exponents, consistent with a Kosterlitz-Thouless transition and the finite size scaling analysis for this transition, have been found.}

\begin{document}
\maketitle
Like charged particles immersed in a homogeneous neutralizing background form crystals at low temperature \cite{Wigner:34,Platzman:74,Monarkha:book:03} ; these crystals are called Wigner crystals after the seminal work of E.P. Wigner in 1934 on electrons in metals \cite{Wigner:34}. Since the original work of Wigner, Coulomb crystals have been observed in a large variety of systems: in plasma physics \cite{Morfill:09,Nosenko:09}, in colloids science \cite{Pertsinidis:01,Levin:02}, in semiconductors \cite{vonKlitzing:80,Tsui:82,Mokashi:12} and in biology\cite{Levin:02}. Two dimensional Wigner crystals are observed in complex plasmas \cite{Morfill:09,Nosenko:09}, in electrons trapped on the surface of liquid Helium\cite{Cole:69,Crandall:71,Grimes:79,Gallet:82}, in laser-cooled $^9$Be$^+$ ions confined in Penning traps \cite{Mitchell:98}, in inversion layer of semiconductors at low temperature \cite{Chaplik:72}.\\
Colloids and dusty plasmas are classical systems ; the classical regime for electrons and ions is defined, in absence of magnetic field, when the Fermi energy is small compared to interaction energy and temperature ; for surface electrons, it corresponds to low surface density \cite{Monarkha:book:03,Waintal:06,Clark:09}.\\
The ground-state of the classical two dimensional Wigner crystal is known to be a triangular lattice \cite{Bonsall:77,Totsuji:78,Antlanger:14b,Platzman:74,Monarkha:book:03} and it is worthwhile to outline that the long ranged nature of the interaction in Coulomb systems does not fulfill the hypothesis of the Mermin theorem on the abscence of long ranged  cristalline order in two dimensions \cite{Mermin:68}.\\
For all experimental systems mentioned above, the study of the phase transition between the ordered Wigner crystal and the disordered fluid-like phases has peculiar importance. Not only for a better understanding of the structural properties of these systems, but also for the theoretical study of the two dimensional meltings \cite{Kosterlitz:74,Nelson:79,Young:79,Nelson:book:83,Strandburg:88,Weber:95,Jaster:98,Bernard:11,Gribova:11,Qi:14,Qi:15,Keim:07,Gasser:10,Deutschlander:14,Lechner:09,Lechner:13,Prestipino:11,Prestipino:14}.\\
In this letter, we report an extensive Monte Carlo study of the melting of the classical two dimensional Wigner crystal. The system is made of $N$ charged point particles confined in a two dimensional plane ; periodic boundary conditions in two dimensions are used. The interaction energy between a pair of particles is the Coulomb energy  $V(r)=Q^2/r$ with $Q$ the charge of particles and $r$ the distance in the plane between both particles. The charges of particles are neutralized by an uniform background of charge density $\sigma_0$ ; electroneutrality of the system reads as $NQ+\sigma_0S=0$ with $S$ the surface of the simulation cell. The number density is noted $\rho=N/S$ and $\sigma_0=-Q\rho$. This system is a One Component Plasma (OCP) confined to a plane.\\
With these notations, for a given configuration $\{\bm{r}_i\}_{1\leq i\leq N}$ of the charged particles, the total energy of the system is given by
\begin{equation}
\label{Total-Energy}
\begin{array}{ll}
\displaystyle E &\displaystyle = \frac{Q^2}{2}\sum_{i=1}^N\sum_{j=1}^N\sum_{\bm{S}_n}\mbox{}'\frac{1}{\mid \bm{r}_i-\bm{r}_j+\bm{S}_n\mid}\\
&\\
&\displaystyle +Q\sigma_0\sum_{i=1}^N\sum_{\bm{S}_n}\mbox{}'\int_{\bm{S}_0}d\bm{r}\frac{1}{\mid \bm{r}_i-\bm{r}+\bm{S}_n\mid}\\
&\\
&\displaystyle + \frac{\sigma_0^2}{2}\sum_{\bm{S}_n}\mbox{}'\int_{\bm{S}_0}d\bm{r}'\int_{\bm{S}_0}d\bm{r}\frac{1}{\mid \bm{r}'-\bm{r}+\bm{S}_n\mid}
\end{array}
\end{equation}
\begin{figure}
\includegraphics[width=3.in]{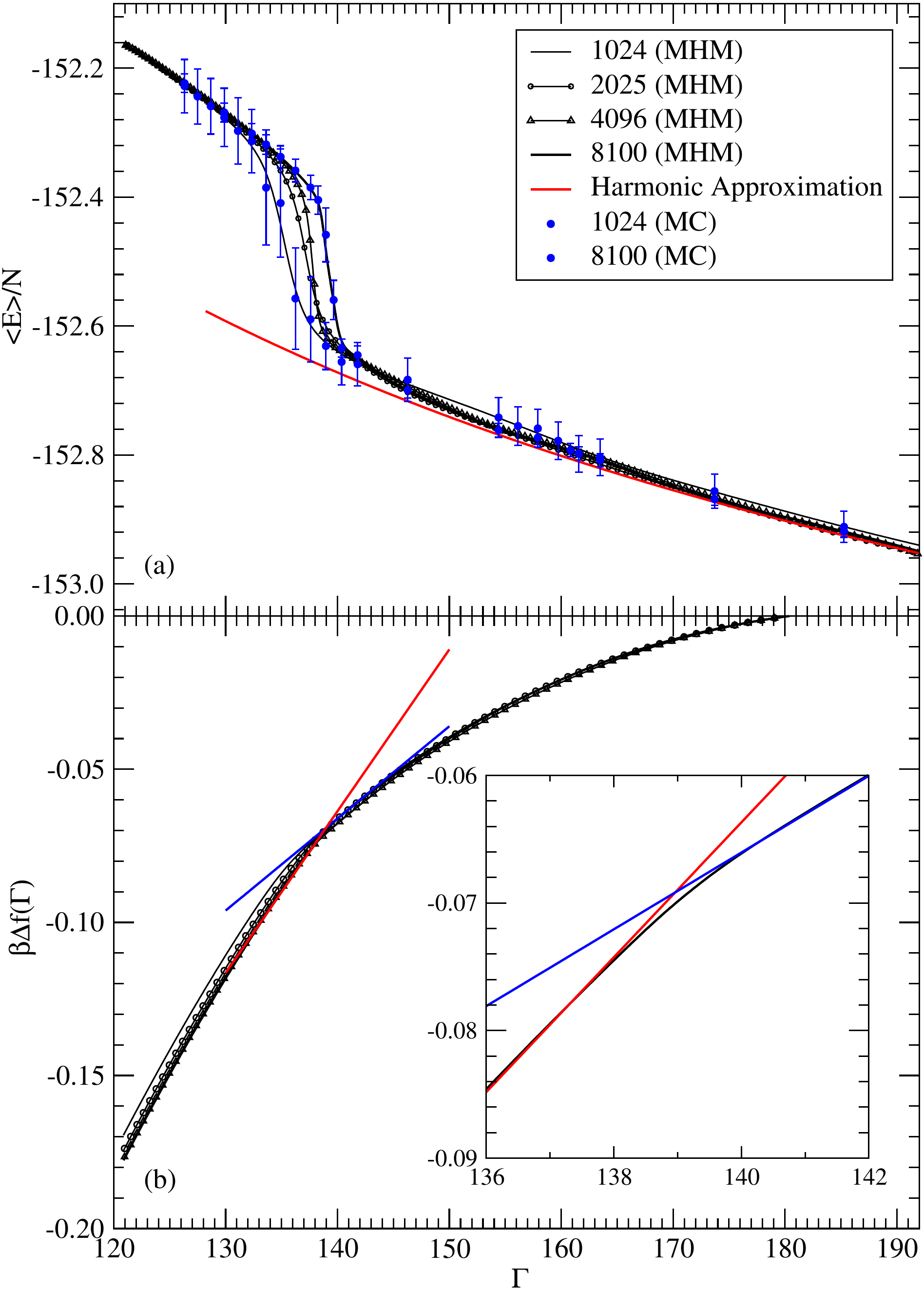}
\caption{Thermodynamical properties of the one component plasma confined in a plane (2D). (a) Excess internal energy per particle as function of the coupling constant $\Gamma$. The symbols in blue with error bars are averages energies computed in Monte-Carlo (MC) simulations. The red line is the internal energy of the 2D Wigner crystal in the low temperature limit as predicted by the harmonic approximation Eq.(\ref{Energy_Solid}). (b) Deviation of the excess free energy per particle from the Wigner crystal free energy as function of $\Gamma$ ; $\beta\Delta f(\Gamma)$ is computed with Eq.(\ref{delta_F}). Inset : enlargement near the phase transition.}
\label{fig1}
\end{figure}
with $\bm{S}_0$ the simulation cell and $\bm{S}_n$, its periodic image. The lattice sums in Eq.\ref{Total-Energy} are computed with the Ewald method\cite{Mazars:11}.\\
The Monte Carlo simulations are done at finite temperature $T$ in the canonical ensemble with a variable shape of $\bm{S}_0$, but at constant surface area. The trial move for the shape of the box with the Ewald method is described in ref.\cite{Weis:01} ; it is particularly well adapted to study solid-solid and solid-liquid transitions and it had been used for the study of the crystal phases of Coulomb \cite{Weis:01} and Yukawa bilayers \cite{Mazars:08}.\\ 
The only relevant thermodynamical variable in the classical regime is the coupling constant $\Gamma=Q^2\sqrt{\pi\rho}/k_B T$.\\
The numerical simulations are performed on systems with $N=1024$, $2025$, $4096$ and $8100$ particles ; for each system size, we have performed MC simulations for about 20 different values of $\Gamma$ ranging from 126 to 214. The analysis of data is done with the multiple histogram method (MHM) \cite{Ferrenberg:88,Newman:book:99,Antlanger:14a,Bhanot:92} and the finite size scaling theory \cite{Challa:86,Binder:87,Peczak:89,Lee:91,Barber:book:83,Privman:book:90,Binder:01,Bhanot:92}.\\
We define a MC-cycle as a trial move of $N$ particles and a trial change of the shape of the simulation box. Equilibrations for each coupling constants are done with $7.5\times 10^5$ MC-cycles for systems with $N=1024$ and $2\times 10^5$ MC-cycles for $N=8100$. In the Monte Carlo sampling of the phase space, after equilibration, the averages are taken over  $2.5\times 10^6$ MC-cycles for systems with $N=1024$ and with $3.0-7.5\times 10^5$ MC-cycles for $N=8100$. For systems size $N=2025$ and $4096$, the numbers of MC-cycles used for equilibrations and averages are in the same range. With the code and the parameters of the Ewald method used in this study, the typical $cpu$-times to perform $2.0\times 10^4$ MC-cycles for the system with $N=8100$ is about 48 hours ; this computing time includes the computational effort for the Voronoi constructions \cite{Voronoi:Book} done every MC-cycle.\\
After equilibration, the trajectories for all systems size and all coupling constants are saved to permit the implementation of the multiple histogram reweighting method (MHM) \cite{Ferrenberg:88,Newman:book:99,Antlanger:14a,Bhanot:92} and the finite size analysis of the phase transition \cite{Challa:86,Binder:87,Peczak:89,Lee:91,Barber:book:83,Privman:book:90,Binder:01,Bhanot:92} ; we define the linear size of systems as $L=\sqrt{N/\rho}$.\\
In refs.\cite{Chen:95,Derzsi:14}, it is shown that to observe a stable hexatic phase one needs simulations long enough such as the phase space of the system is correctly sampled ; this is related to the critical properties of hexatic phases \cite{Nelson:book:83}. For Lennard-Jones systems, the stability of the hexatic phase is confirmed by longer computations \cite{Wierschem:11}.\\ 
An estimate of the sampling of the phase space is given by the average length of the trajectory of particles. In the Molecular Dynamics reported in refs.\cite{Chen:95,Derzsi:14}, we may estimate the average length of trajectories from the average velocity and the total duration of the computations. From the data reported in these works, we found that the lengths of trajectories are 25-70 $L$.  In the MC reported in the present work, with an amplitude of the trial moves as $0.004-0.006$ $L$, the average length of trajectories are about 550 $L$ for equilibration and 2000 $L$ for averages in systems with $N=1024$, and about 130 $L$ for equilibration and 440 $L$ for averages in systems with $N=8100$.\\
In the following, we note $<.>$ the averages obtained with the MHM and by $<.>_ {MC}$ the averages computed in Monte Carlo simulations.\\
Figure \ref{fig1} gives the thermodynamical properties of the OCP confined in a plane. The excess internal energy per particles is computed as 
\begin{equation}
\label{Energy}
\displaystyle <E>=-\frac{\partial}{\partial \beta} \ln Z_N(\Gamma)
\end{equation}
where $\ln Z_N(\Gamma)$ is the partition function computed with MHM. The pressure is related to $<E>$ and the contribution to energy due to the neutralizing background (see e.g. refs. \cite{Hoover:71,Navet:80}.)\\
For large value of $\Gamma$, the low temperature limit, the system form a triangular Wigner crystal, the harmonic approximation allows to represent the energy in this limit as \cite{Ashcroft:book:76,Travesset:14}
\begin{equation}
\label{Energy_Solid}
\displaystyle \frac{<E>}{N Q^2\sqrt{\rho}}\simeq e_0+C_0\frac{1}{\Gamma}
\end{equation}
Letting $e_0$ as a free parameter, we find that it does not depend on the number of particles and we have $e_0=-1.961(1)$, this value is very close to the known Madelung constant ($c_M= -1.960515788...$) of the triangular lattice\cite{Bonsall:77,Antlanger:14b}.\\
If we fix $e_0=c_M$, then we obtain, for all system sizes, $C_0=1.845(5)$. In Fig.\ref{fig1} (a), we represent  Eq.(\ref{Energy_Solid}) with $e_0=c_M$ and $C_0=1.845(5)$ as a thick solid red line.\\
 Eq.(\ref{Energy}), with the MHM, permits to obtain the ground state energy of a model system  to a very good accuracy \cite{Antlanger:14a} ; therefore, it provides a reference point for the computation of the free energy.\\ 
The partition function $Z_{s}(\Gamma)$ in the low temperature limit is obtained by integration of Eq.(\ref{Energy_Solid}) and, with the definition given in Eq.(\ref{Energy}), we find
\begin{equation}
\label{F_Solid}
\displaystyle \frac{1}{N}\ln Z_{s}(\Gamma)= Z_0-\left(\frac{2\rho}{\pi \sqrt{3}}\right)^{1/2}\left(c_M\Gamma + C_0\ln\Gamma\right)
\end{equation}
where $Z_0$ is a constant of integration and the prefactor $(2\rho/\pi \sqrt{3})^{1/2}$ stems from the definition of $\Gamma$ and the geometric features of the triangular lattice \cite{Antlanger:14b,Travesset:14}.\\
The deviation of the excess free energy per particle from the Wigner crystal free energy as function of $\Gamma$ is represented in Fig.\ref{fig1} (b) ; it is computed as 
\begin{figure}
\onefigure[width=3.in]{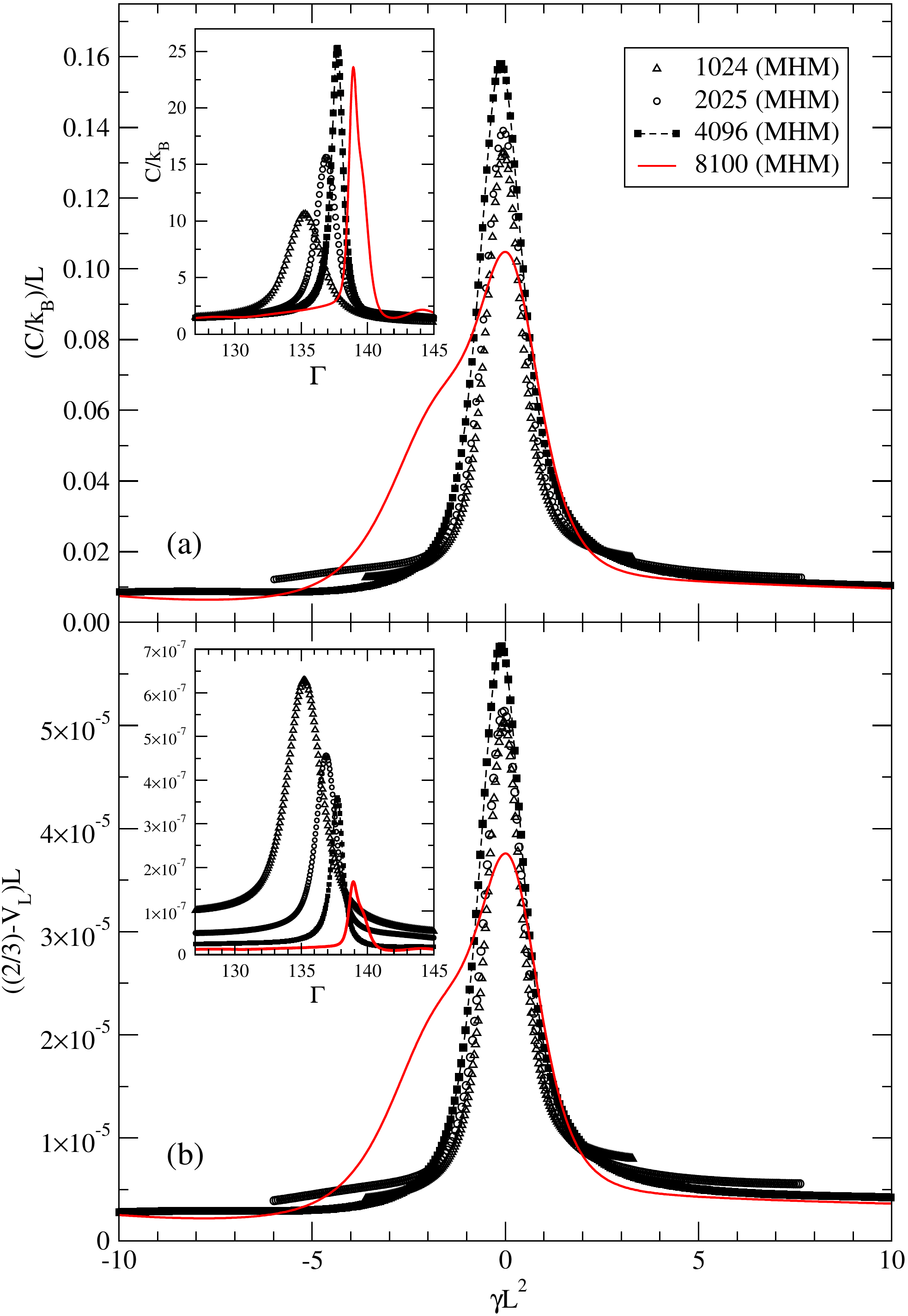}
\caption{Finite-size scaling of the thermodynamical observables for the one component plasma confined in a plane (2D). All data represented are computed with the multiple histogram method (MHM). (a) Scaling of the specific-heat $C/k_B$ ; inset : specific-heat as function of coupling constant $\Gamma$ for different sizes. (b) Scaling of the fourth-order cumulants of the energy ; inset : cumulants as function of $\Gamma$.}
\label{fig2}
\end{figure}
\begin{equation}
\label{delta_F}
\displaystyle \beta\Delta f(\Gamma)= \frac{1}{N}\ln Z_{N}(\Gamma)-\frac{1}{N}\ln Z_{s}(\Gamma)
\end{equation}
where $\ln Z_N(\Gamma)$ is the partition function obtained with MHM.\\
The slope discontinuity of the tangents near the phase transition is an indication of a first order phase transition.\\  
Since the only relevant thermodynamical parameter is $\Gamma$, the first order transition is driven by the change of the temperature \cite{Challa:86,Binder:87}.\\
The specific heat is defined by
\begin{equation}
\label{specific_heat}
\displaystyle \frac{C}{k_B}=\frac{1}{N}\left(\frac{\Gamma}{Q^2\sqrt{\pi\rho}}\right)^2\left(<E^2>-<E>^2\right)
\end{equation}
The kurtosis \cite{Bhanot:92} or the Binder's fourth cumulant \cite{Lee:91} is given by 
\begin{equation} 
\label{Cumulant_E}
\displaystyle V_L=1-\frac{<E^4>}{3(<E^2>)^2} 
\end{equation}
The variations of $V_L$ with $\Gamma$ for different system sizes is consistent with a first oder phase transition \cite{Challa:86,Lee:91,Bhanot:92} ; more precisely, the very small deviation of $V_L$ from 2/3 indicate a weak first order transition \cite{Bhanot:92}. The extremums of $C/k_B$ and $V_L$ give $\Gamma_c(L)$, the rounding coupling constant due to the finite size of the simulated systems. We define the reduced temperature as 
\begin{equation}
\label{gamma}
\displaystyle \gamma=\frac{1}{\Gamma}-\frac{1}{\Gamma_c(L)}
\end{equation}
and in Fig.\ref{fig2} we represent the collpase of the observables $C/k_B$ (a) and $V_L$ (b) as functions of $\gamma L^2$.\\
The scaling with $\gamma L^2$ is excellent, however the amplitude of the thermodynamical observables scales with $L$ and not with $L^2$ as excepted for a strong first order phase transition \cite{Challa:86,Binder:87,Lee:91}. We interpret this anomalous scaling of the thermodynamical observables as induced by the weakness of the first order phase transition between the liquid and hexatic phases \cite{Peczak:89}. A similar deviation to the finite size scaling of first order transition is observed in five-state Potts model in two dimensions \cite{Peczak:89} ; the pseudocritical behavior stems from a finite, but very large ($>L$), correlation length (see e.g. Fig.16 in ref.\cite{Peczak:89}).\\
The data for $N=8100$ deviates significantly from the scaling functions. Periodic boundary conditions favor the crystal phases. In small systems ($N=1024$ and 2025), the range of stability of the hexatic phase with short ranged positional order and long ranged orientational order is artificially reduced because of the positional order induced by the periodic boundary conditions.\\ 
In multiple histograms method, one is able to compute the partition function $\ln Z_N(\Gamma)$ for any values of $\Gamma$ in the range covered by the MC simulations \cite{Ferrenberg:88,Bhanot:92,Newman:book:99} and to interpolate the variations of observables with $\Gamma$.\\
In figure \ref{fig3}(a), we represent the histograms of energy computed from the Monte-Carlo trajectories and histograms computed with the MHM for $N=8100$ ; in figure \ref{fig3}(b), we plot the MC-trajectories of the total energy used to construct the MC-Histograms in (a). The orange trajectory and histogram are for the hexatic phase very close to the hexatic-liquid transition and, the greens are for the hexatic phase very close to the solid-hexatic transition.\\
For all systems sizes, we may represent the probability distribution of the total energy as a superposition of gaussian distributions as \cite{Challa:86,Lee:91,Binder:87}
\begin{figure}
\onefigure[width=3.in]{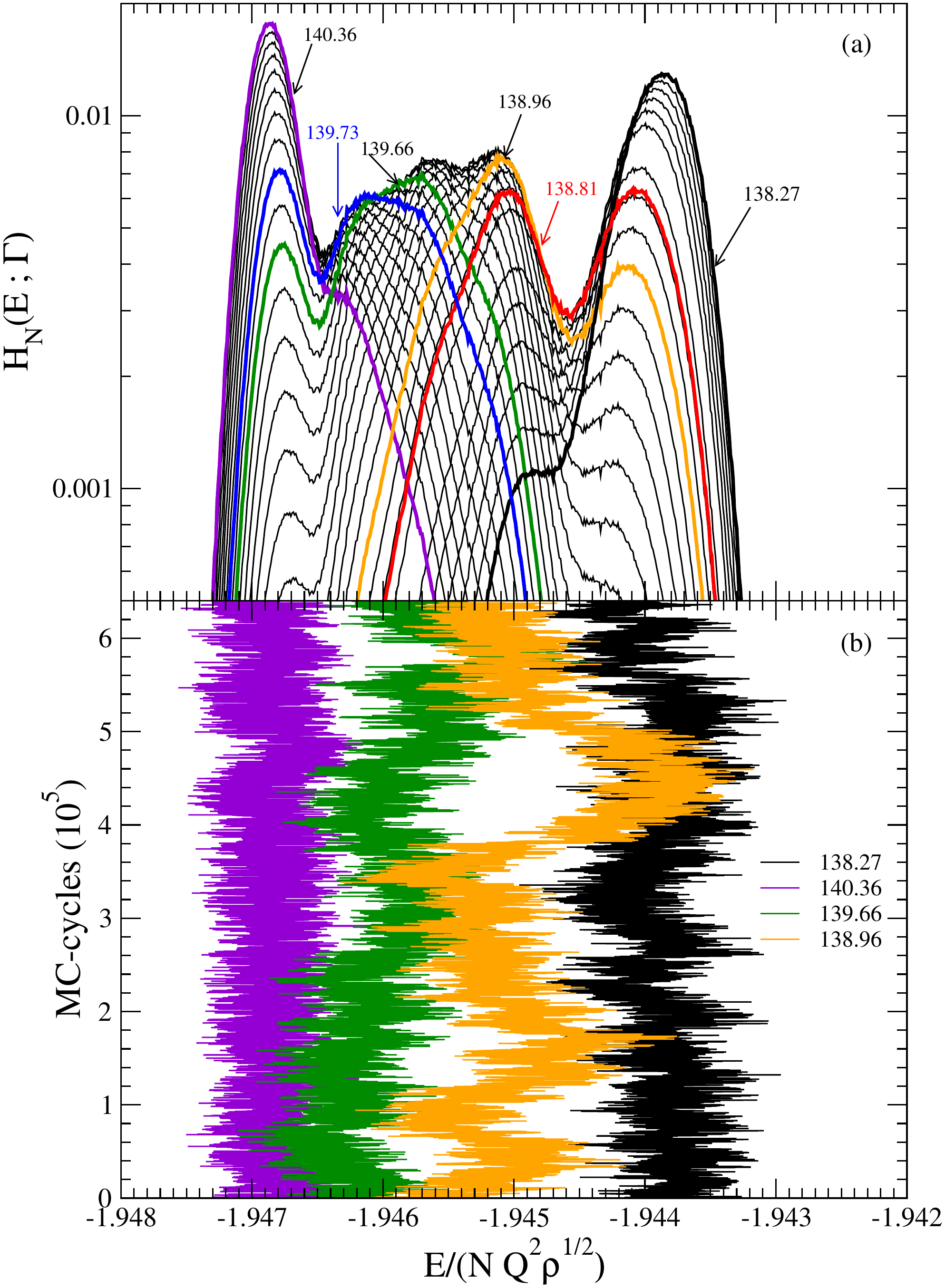}
\caption{Histograms (a) and trajectories (b) of the energy close to the phase transitions for systems with $N=8100$. The thin black lines are histograms interpolated from the MC-histograms with the multiple histogram method. The thick red and blue lines are histograms of energies very close to the transitions, used to estimate $\Gamma_h$ and $\Gamma_s$. The colored histograms represented in (a) with thick lines are MC-histograms computed from the energy trajectories shown in (b) with the same color.} 
\label{fig3}
\end{figure}
\begin{equation}
\label{gaussian}
\displaystyle H_N(E;\Gamma)=\sum_{a} H_a(\Gamma)\exp\left(-\frac{(E-E_a(\Gamma))^2}{2\sigma_a^2(\Gamma)}\right)
\end{equation}
with $a\in\{s,h,l\}$ (triple peaks) or $a\in\{s,l\}$ (double peaks)  where $s$ stands for the Wigner crystal phase, $h$ the hexatic phase and $l$ the disordered liquid phase.\\ 
In the plot of histograms, the various phases are represented by the peaks and, for a given size, we locate the temperature of transition between two phases, $1/\Gamma_c(L)$, by requiring that the height of the peaks to be the same.  Assuming that the specific heats do not vary with the temperature close to the phase transitions, the gaussians are centered at $E_a(\Gamma)=E_a+C_a \gamma$.\\
The values for $E_a(\Gamma)$ with $a\in\{s,h,l\}$ (triple peaks) and $a\in\{s,l\}$ (double peaks) are represented on Fig.\ref{fig4} with the excess internal energy computed with MHM by Eq.(\ref{Energy}) for $N=4096$ and 8100. These data permit to locate the transition coupling constants and the range of stability of the hexatic phase, reported in Table \ref{tab1}.\\ 
In Fig.\ref{fig4}, we report also the structure factor $S(\bm{k})$ for the systems with $N=8100$, for the fluid phase ($\Gamma=137.58$), the hexatic phase ($\Gamma=138.96$) and the Wigner crystal ($\Gamma=140.36$). In experiments, the structure factors are obtained from diffraction patterns \cite{Chao:04}; in simulations, $S(\bm{k})$ are computed as 
\begin{figure}
\onefigure[width=3.4in]{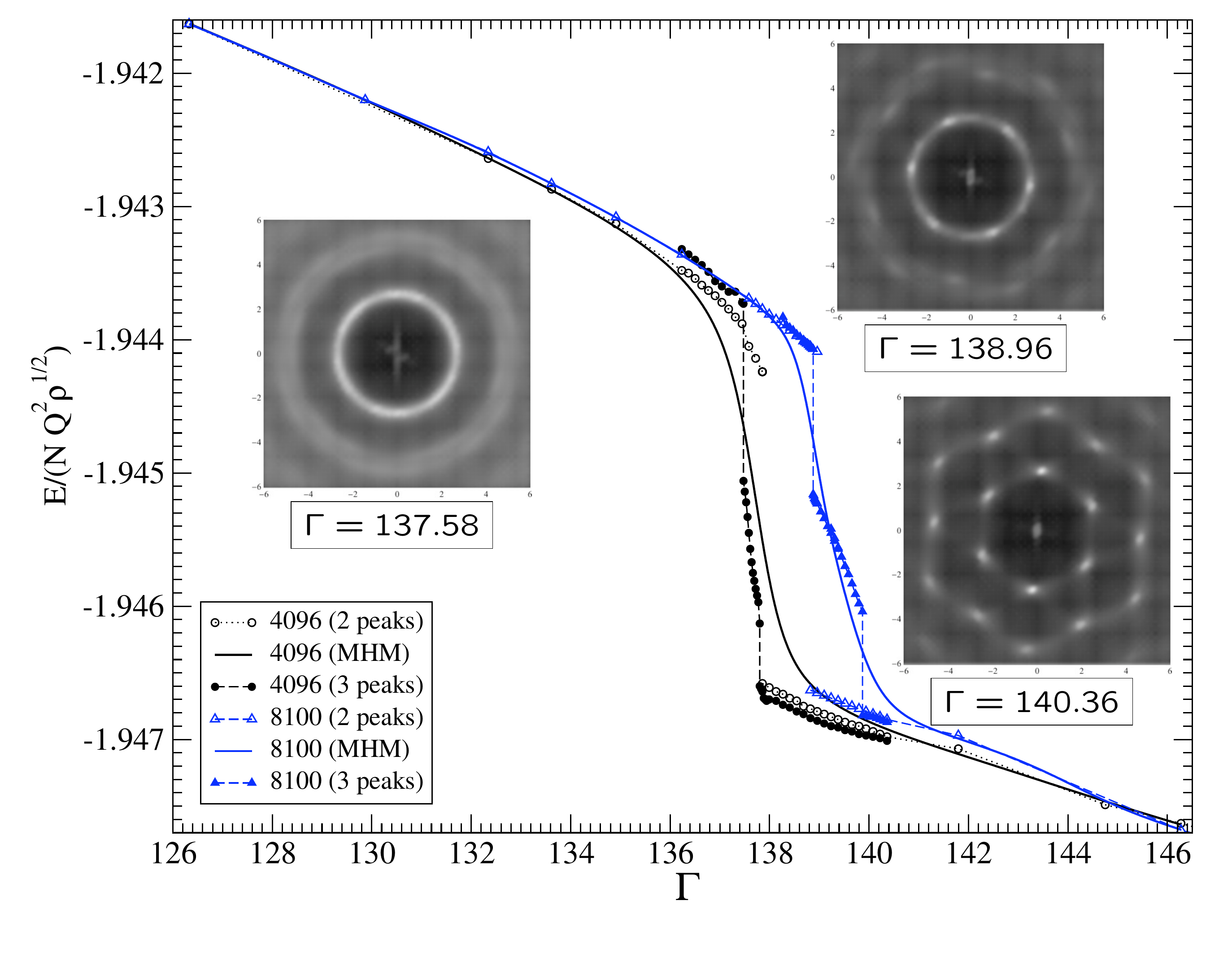}
\caption{Location of the maximums of the double (2 peaks) and triple (3 peaks) peaked  histograms compared to excess internal energies computed with the multiple histogram method  ($N=4096,8100$). The structure factors $S(\bm{k})$ are computed in Monte-Carlo simulations in systems with $N=8100$ ; they are represented as density plots for three values of $\Gamma$.}
\label{fig4}
\end{figure}
\begin{equation}
\label{SF_1}
\displaystyle S(\bm{k})=\frac{1}{N}\left<\rho(\bm{k})\rho(-\bm{k})\right>_{MC}
\end{equation}
with
\begin{equation}
\label{SF_2}
\displaystyle \rho(\bm{k}) =\int d\bm{s}\exp(-i \bm{k}.\bm{r})\rho(\bm{r})=\sum_{n=1}^N\exp(-i \bm{k}.\bm{r}_n)
\end{equation}
The structure factors, shown on Fig.\ref{fig4}, support the identification of the three peaks with the three phases : liquid, hexatic and solid.\\
The bond orientational order parameter $\phi_6$ is a suitable quantity for the study of the melting of triangular lattices. We compute $\phi_6$, the susceptibility $\chi_6$ and the fourth-order cumulant $U_6$ with Voronoi constructions, as described in ref.\cite{Mazars:08}.\\  
\begin{table}
\caption{Estimate of the transition coupling constants and latent heat for the melting of the two dimensional Wigner crystal. $\Gamma_h(L)$ is the coupling constant for the fluid/hexatic transition ; $L_{l/h}$ the latent heat of the fluid/hexatic transition ; $\Delta\Gamma$ and $\Gamma_s(L)$ are respectively crude estimates of the stability range of the hexatic phase and the hexatic/solid transition.}
\label{tab1}
\begin{center}
\begin{tabular}{c|cccc}
\hline
$N$  & $\Gamma_h(L)$ & $L_{l/h}$ & $\Delta\Gamma$ & $\Gamma_s(L)$ \\
\hline
1024 & 135.0$\pm 0.14$ & $1.2\times10^{-3}$ & 0.16 & 135 \\
2025 & 136.7$\pm 0.12$ &$1.2\times10^{-3}$ & 0.19 & 137 \\
4096 & 137.5$\pm 0.1$ & $1.4\times10^{-3}$ & 0.32 & 138 \\
8100 & 138.81$\pm 0.05$ & $1.0\times10^{-3}$ & 1.05 & 140 \\	
\hline
\end{tabular}
\end{center}
\end{table}
\begin{figure}
\onefigure[width=3.in]{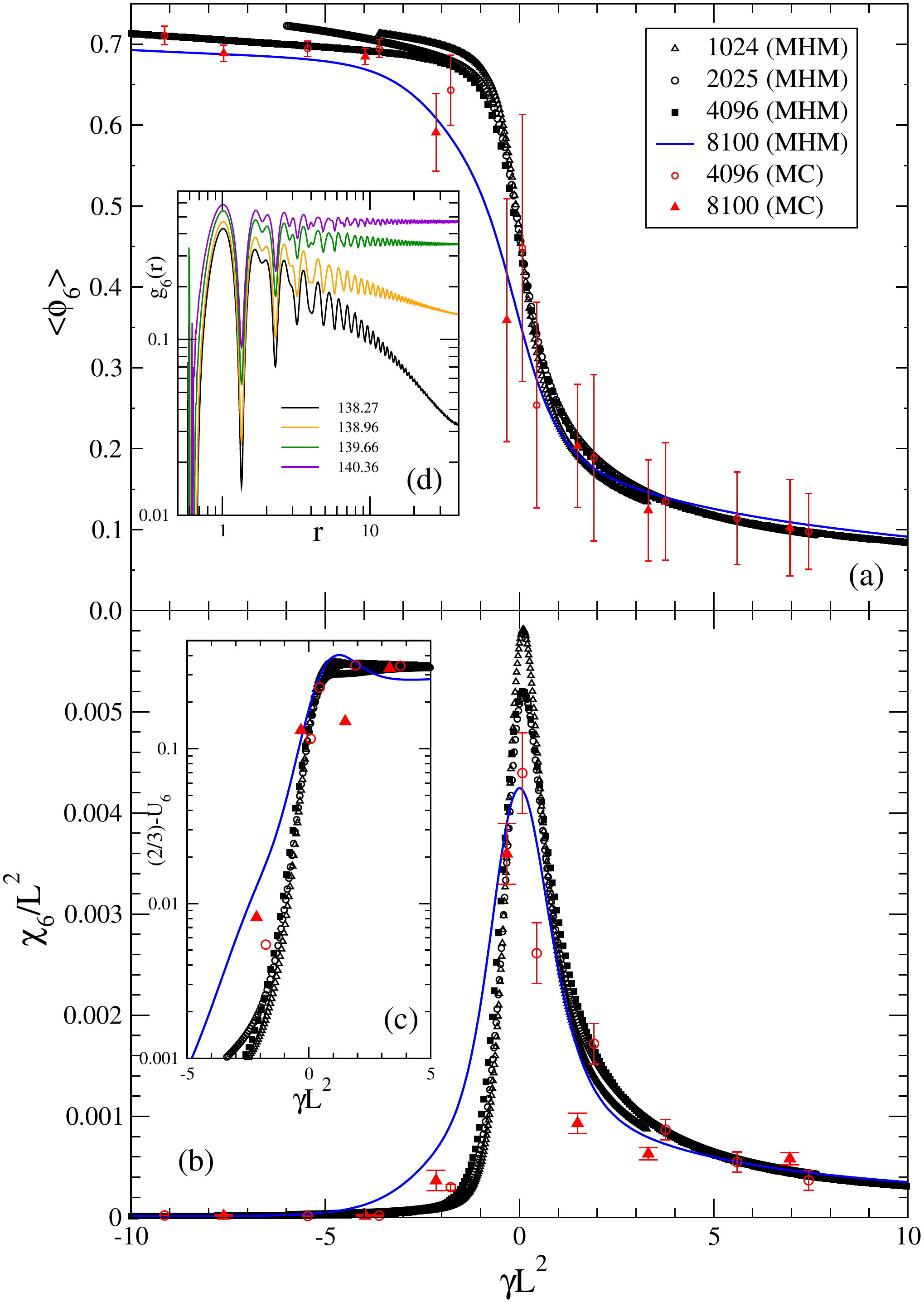}
\caption{Finite-size scaling for the bond orientational  order parameters and derived quantities. (a) Scaling of the bond orientational order parameter $\phi_6$. (b) Scaling of the susceptibility $\chi_6$. (c) Scaling of the fourth-order cumulant of the order parameter $U_6$.(d) Bond orientational correlation functions $g_6(r)$ for the MC-trajectories of Fig.\ref{fig3}(b).}
\label{fig5}
\end{figure}
On Fig.\ref{fig5}, we represent the data collapses of $<\phi_6>$, $\chi_6$ and $U_6$ as functions of the reduced temperature $\gamma L^2$. The finite size scaling depends on the surface of the system (or number of particles) that is consistent with a first order phase transition. As for the thermodynamical observables, data for systems $N=8100$ deviate significantly from the scaling functions ; we interpret this deviation as resulting of an increase of the thermal stability of the hexatic phase in larger systems. On the contrary to the scaling of the amplitude of $C/k_B$ with the system size $L$ observed in Fig.\ref{fig2}, the amplitude of $\chi_6$ scales with $L^2$ as predicted for a first order transition.\\
On Fig.\ref{fig5}(d), we represent the bond orientational correlation functions $g_6(r)$ for the MC-trajectories of Fig.\ref{fig3}(b).\\
In the KTHNY  theory of the two dimensional melting \cite{Kosterlitz:74,Nelson:79,Young:79,Nelson:book:83,Strandburg:88}, the finite size scaling of $\chi_6$ at the liquid-hexatic transition ($\gamma >0$) is governed by $\chi_6 \sim\xi_6^{(2-\eta_6)}$ with $\xi_6$ the correlation length of the order parameter ; it exhibits an essential singularity scaling as $\exp(b/\gamma^{\nu})$ at the transition \cite{Kosterlitz:74}. In the present study, no set of critical exponents, consistent with the KTHNY theory or with the XY-model \cite{Kosterlitz:74}, have been found to achieve data collapses better, or at least as good as, thoses done for a first order phase transition \cite{Barber:book:83,Privman:book:90,Binder:01}.\\
All the results reported in Figs.\ref{fig1}-\ref{fig5} support a weak first order phase transition for the liquid-hexatic transition in the melting of the Wigner crystal.\\
Based on the analysis of the triple peaked histograms, we report in Table \ref{tab1} the transition coupling constants for the hexatic/solid transition $\Gamma_s(L)$ and for the weak first order liquid/hexatic transition $\Gamma_h(L)$, found for all systems size. The errors bars on values of $\Gamma_h(L)$ are obtained by neighboring histograms with peaks about the same height. The latent heats of the liquid/hexatic transition are computed as $L_{l/h}=E_l-E_h$ (see also Fig.\ref{fig4}) ; the uncertainties on the numerical values of $L_{l/h}$ are estimated about 10\%. The small values of the latent heat found for the fluid/hexatic transition is another signature of the weakness of the first order phase transition ; the same order of magnitude is found for hard disks systems \cite{Qi:15}.\\
The fluid/hexatic phase transition for the Coulomb system studied in the present work is compatible with the grain boundary induced melting found in experiments on complex plasmas \cite{Nosenko:09}. This transition has close similarities with the melting of hard disk systems \cite{Bernard:11,Weber:95,Jaster:98} and hard spheres in slab geometry \cite{Qi:14}, but it is clearly different from the KTHNY mechanism found for superparamagnetic colloids at air-water interface in an external magnetic field \cite{Keim:07,Gasser:10,Deutschlander:14,Lechner:09,Lechner:13}.\\
As explained before, the stability of the Wigner crystal is increased because of the periodic boundary conditions (see also ref.\cite{Qi:15}) ; nevertheless, the detailled analysis of the histograms of energy (Figs.\ref{fig3} and \ref{fig4}) have permitted to locate approximatively the hexatic/solid transition ($\Gamma_s(L)$ in Table \ref{tab1}). Figures \ref{fig3} and \ref{fig4} tend to support a first order phase transition for the hexatic/solid transition, in agreement with a previous study done by B.K. Clark and co-workers on 2D quantum Coulomb systems \cite{Clark:09}. However, with the system sizes considered in the present work, we haven't yet been able to reach a definitive conclusion on the nature and order of the hexatic/solid transition. 
\acknowledgments
This work is supported by \emph{Th\`eme 2 - LabEx PALM} (projet - SLAB - ANR-10-LABX-0039-PALM) and project ECOS-Nord C14P01. The author acknowledges the computation facilities (iDataPlex - IBM) provided by  {\it Direction Informatique\/} of {\it Universit\'e Paris-Sud\/}.


\begin{thebibliography}{0}
  
  \bibitem{Wigner:34}
  \Name{Wigner E.P.}
  \REVIEW{Phys. Rev.}{46}{1934}{1002}.
  
  \bibitem{Platzman:74}  
  \Name{Platzman P.M. \and Fukuyama H.}
  \REVIEW{Phys. Rev. B}{10}{1974}{3150} 
  
   \bibitem{Monarkha:book:03}
  \Name{Monarkha Y. \and Kono K.}
  \Book{Two-Dimensional Coulomb Liquids and Solids - Springer Series in Solid-State Science.}
  \Vol{142}
  \Publ{Springer, Berlin}
  \Year{2004}.

 \bibitem{Morfill:09}
  \Name{Morfill G. \and Ivlev A.}
  \REVIEW{Rev. Mod. Phys.}{81}{2009}{1353}.

  \bibitem{Nosenko:09}
  \Name{Nosenko V., Zhdanov S.K., Ivlev A.V., Knapek C.A. \and Morfill G.E.}
  \REVIEW{Phys. Rev. Lett.}{103}{2009}{015001}.

  \bibitem{Pertsinidis:01}   
  \Name{Pertsinidis A. \and Ling X.S.}
  \REVIEW{Phys. Rev. Lett.}{87}{2001}{098303}.
  
  \bibitem{Levin:02}
  \Name{Levin Y.}
  \REVIEW{Rep. Prog. Phys.}{65}{2002}{1577}.
  
  \bibitem{vonKlitzing:80}   
  \Name{von Klitzing K., Dorba G. \and Pepper M.}
  \REVIEW{Phys. Rev. Lett.}{45}{1980}{494}.
  
  \bibitem{Tsui:82}   
  \Name{Tsui D. C., Stormer H.L. \and Gossard A.C.}
  \REVIEW{Phys. Rev. Lett.}{48}{1982}{1559}.
  
  \bibitem{Mokashi:12}
  \Name{Mokashi A., Li S., Wen, B. Kravchenko S.V., Shashkin A.A., Dolgopolov, V.T. \and Sarachik}
  \REVIEW{Phys. Rev. Lett.}{109}{2012}{096405}.

   \bibitem{Cole:69}
  \Name{Cole M.W. \and Cohen M.H.}
  \REVIEW{Phys. Rev. Lett.}{23}{1969}{1238}.
  
  \bibitem{Crandall:71}
  \Name{Crandall R.S. \and Williams R.}
  \REVIEW{Phys. Lett. A}{34}{1971}{404}.
  
   \bibitem{Grimes:79}
  \Name{Grimes C.C. \and Adams G.}
  \REVIEW{Phys. Rev. Lett.}{42}{1979}{795}.
  
  \bibitem{Gallet:82} 
  \Name{Gallet F., Deville G., Vald\`es A. \and Williams F.I.B.}
  \REVIEW{Phys. Rev. Lett.}{49}{1982}{212}.
  
  \bibitem{Mitchell:98} 
  \Name{Mitchell T.B., J.J. Bollinger J.J., Dubin D.H.E.,  \etal}
  \REVIEW{Science}{282}{1998}{1290} ; \REVIEW{Phys. Plasmas} {6}{1999}{1751}.
  
  \bibitem{Chaplik:72}
  \Name{Chaplik A.V.}
  \REVIEW{Sov. Phys JETP}{35}{1972}{395}. 
  
  \bibitem{Waintal:06}
  \Name{Waintal X.}
  \REVIEW{Phys. Rev. B}{73}{2006}{075417}. 
 
  \bibitem{Clark:09}
  \Name{Clark B.K., Casula M. \and Ceperley D.M.}
  \REVIEW{Phys. Rev. Lett.}{103}{2009}{055701}. 
  
  \bibitem{Bonsall:77}
  \Name{Bonsall L. \and Maradudin A.A.}
  \REVIEW{Phys. Rev. B}{15}{1977}{1959}.

  \bibitem{Totsuji:78}
  \Name{Totsuji H.}
  \REVIEW{Phys. Rev. A}{17}{1978}{399}.

  \bibitem{Antlanger:14b}
  \Name{Antlanger M., Mazars M., \v{S}amaj L., Kahl G. \and Trizac E.}
  \REVIEW{Mol. Phys.}{112}{2014}{1336}.


  \bibitem{Mermin:68}
  \Name{Mermin N.D.}
  \REVIEW{Phys. Rev.}{176}{1968}{250}.


  \bibitem{Kosterlitz:74}
  \Name{Kosterlitz J.M.}
  \REVIEW{J. Phys. C: Solid State Phys.}{7}{1974}{1046}.

  \bibitem{Nelson:79}
  \Name{Nelson D.R. \and Halperin B.I.}
  \REVIEW{Phys. Rev. B}{19}{1979}{2457}.
  
   \bibitem{Young:79}
  \Name{Young A.P.}
  \REVIEW{Phys. Rev. B}{19}{1979}{1855}.
    
  \bibitem{Nelson:book:83}
  \Name{Nelson D.R.}
  \Book{Defect-mediated Phase Transitions - Phase Transitions and Critical Phenomena.}
  \Editor{Domb C. \and Lebowitz J.L.}
  \Vol{7}
  \Publ{Academic Press, London}
  \Year{1983}.

  \bibitem{Strandburg:88}
  \Name{Strandburg K.J.}
  \REVIEW{Rev. Mod. Phys.}{60}{1988}{161}. 
  
  
  \bibitem{Weber:95}
  \Name{Weber H., Marx D. \and Binder K.}
  \REVIEW{Phys. Rev. B}{51}{1995}{14636}.
  
   \bibitem{Jaster:98}
  \Name{Jaster A.}
  \REVIEW{EPL}{42}{1998}{277} ;  \REVIEW{Phys. Rev. E}{59}{1999}{2594}. 
  
  \bibitem{Bernard:11}
  \Name{Bernard E.P. \and Krauth W.}
  \REVIEW{Phys. Rev. Lett.}{107}{2011}{155704}.
  
  \bibitem{Gribova:11}
  \Name{Gribova N., Arnold A., Schilling T. \and Holm C.}
  \REVIEW{J. Chem. Phys.}{135}{2011}{054514}.
  
  \bibitem{Qi:14}
  \Name{Qi W., Gantapara A.P. \and Dijkstra M.}
  \REVIEW{Soft Matter}{10}{2014}{5449}.
\bibitem{Qi:15}\Name{Qi W. \and Dijkstra M.}\REVIEW{ Soft Matter}{DOI: 10.1039/C4SM02876G}{2015}{.}
  

  \bibitem{Keim:07}
  \Name{Keim P., Maret G., \and von Gr\"unberg H.H.}
  \REVIEW{Phys. Rev. E}{75}{2007}{031402}.
  
  \bibitem{Gasser:10}
  \Name{Gasser U., Eisenmann C., Maret G. \and Keim P.}
  \REVIEW{Chem. Phys. Chem.}{11}{2010}{963}.
  
  \bibitem{Deutschlander:14}
  \Name{Deutschl\"{a}nder S., Puertas A.M., Maret G. \and Keim P.}
  \REVIEW{Phys. Rev. Lett.}{113}{2014}{127801}.
  
  \bibitem{Lechner:09}
  \Name{Lechner W.\and Dellago C.}
  \REVIEW{Soft Matter}{5}{2009}{2752}.

  \bibitem{Lechner:13}
  \Name{Lechner W., Polster D., Maret G., Keim P. \and Dellago C.}
  \REVIEW{Phys. Rev. E}{88}{2013}{060402(R)}.    

  
  
  \bibitem{Prestipino:11}
  \Name{Prestipino S.,  Saija F. \and Giaquinta, P.V.}
  \REVIEW{Phys. Rev. Lett.}{106}{2011}{235701}.
  
  \bibitem{Prestipino:14}
  \Name{Prestipino S. \and Saija F.}
  \REVIEW{J. Chem. Phys.}{141}{2014}{184502}.
  

 \bibitem{Mazars:11}
  \Name{Mazars M.}
  \REVIEW{Phys. Rep.}{500}{2011}{43}.
\bibitem{Weis:01}\Name{Weis J.-J., Levesque D. \and Jorge S.} \REVIEW{Phys. Rev. B}{63}{2001}{045308}.
\bibitem{Mazars:08}\Name{Mazars M.} \REVIEW{EPL}{84}{2008}{55002}.
  
  \bibitem{Ferrenberg:88}
  \Name{Ferrenberg A.M. \and Swendsen R.H.}
  \REVIEW{Phys. Rev. Lett.,}{61}{1988}{2635} ; \REVIEW{Phys. Rev. Lett.,}{63}{1989}{1135}.

  \bibitem{Newman:book:99}
  \Name{Newman M.E.J. \and  Barkema G.T.}
  \Book{Monte Carlo Methods in Statistical Physics.}
  \Publ{Oxford University Press, Oxford}
  \Year{1999}.

  \bibitem{Antlanger:14a}
  \Name{Antlanger M., Doppelbauer G., Mazars M. \and Kahl G. }
  \REVIEW{J. Chem. Phys.}{140}{2014}{044507}.
  
    \bibitem{Bhanot:92}
  \Name{Bhanot G., Lippert T., Schilling K. \and Ueberholz P.}
  \REVIEW{Nucl. Phys. B}{378}{1992}{633}.
  
  \bibitem{Challa:86}
  \Name{Challa M.S.S., Landau, D.P. \and Binder K.}
  \REVIEW{Phys. Rev. B}{34}{1986}{1841}.
  
  \bibitem{Binder:87}
  \Name{Binder K.}
  \REVIEW{Rep Prog. Phys.}{50}{1987}{783}.
   
   \bibitem{Peczak:89}
  \Name{Peczak P. \and Landau D.P.}
  \REVIEW{Phys. Rev. B}{39}{1989}{11932}.
   
  \bibitem{Lee:91}
  \Name{Lee J. \and Kosterlitz M.}
  \REVIEW{Phys. Rev. B}{43}{1991}{3265}.

  \bibitem{Barber:book:83}
  \Name{Barder M.N.}
  \Book{Finite-size Scaling - Phase Transitions and Critical Phenomena.}
  \Editor{Domb C. \and Lebowitz J.L.}
  \Vol{8}
  \Publ{Academic Press, London}
  \Year{1983}.

  \bibitem{Privman:book:90}
  \Name{Privman V.}
  \Book{Finite Size Scaling and Numerical Simulation.}
  \Publ{World Scientific, Singapore}
  \Year{1990}.

  \bibitem{Binder:01}
  \Name{Binder K. \and Luijten E.}
  \REVIEW{Phys. Rep.}{344}{2001}{179}.
  
  \bibitem{Voronoi:Book}
  \Name{Okabe A., Boots B., Sugihara K. \and Nok Chiu S.}
  \Book{Spatial Tessellations: Concepts and Applications of Voronoi Diagrams (Second Edition).}
  \Publ{John Wiley \& Sons Inc., New York}
  \Year{2000}.
\bibitem{Chen:95} \Name{Chen K., Kaplan T. \and Mostoller M.}  \REVIEW{Phys. Rev. Lett.}{74}{1995}{4019}.
\bibitem{Derzsi:14}\Name{Derzi A., Kov\'acs A., Donk\'o Z \and Hartmann P.}\REVIEW{Phys. Plasma}{21}{2014}{023706}.
\bibitem{Wierschem:11} \Name{Wierschem K. \and Manousakis E.} \REVIEW{Phys. Rev. B}{83}{2011}{214108}.

  \bibitem{Hoover:71}
  \Name{Hoover W.G., Gray S.G. \and Johnson K.W.}
  \REVIEW{J. Chem. Phys.}{55}{1971}{1128} ; {{\bf 56} }{(1972) }{2207}.

 \bibitem{Navet:80}
  \Name{Navet M., Jamin E. \and Feix M.R.}
  \REVIEW{J. Physique -Lettres}{41}{1980}{L69}.
  
  \bibitem{Travesset:14}
  \Name{Travesset A.}
  \REVIEW{J. Chem. Phys.}{141}{2014}{164501}.

  \bibitem{Ashcroft:book:76}
  \Name{Ashcroft N.W. \and Mermin N.D.}
  \Book{Solid State Physics.}
  \Publ{Brooks/Cole, Thomson Learning, Inc.}
  \Year{1976}.
\bibitem{Chao:04}\Name{Chao C.-Y., Hsu M.-T., Hsieh W.-J., Ho J. T. \and Lin J.B.} \REVIEW{Phys. Rev. Lett.}{93}{2004}{247801}.
  
  
\end{thebibliography}
\end{document}